\documentclass[journal=jacsat,manuscript=article]{achemso}

\usepackage{chemformula} % Formula subscripts using \ch{}
\usepackage[T1]{fontenc} % Use modern font encodings

\author{Giuseppe Antonacci}
\affiliation[Specto]
{Specto Photonics, Via Caradosso 12, 20123 Milan, Italy}
\email{giuseppe@spectophotonics.com}

\author{Kareem Elsayad}
\affiliation[Vienna]
{Division of Anatomy, Center for Anatomy and Cell Biology, Medical University of Vienna, Waehringerstrasse 13, Vienna, A-1090}

\author{Dario Polli}
\affiliation[Specto]
{Specto Photonics, Via Caradosso 12, 20123 Milan, Italy}
\altaffiliation{Dipartimento di Fisica, Politecnico di Milano, Piazza L. da Vinci 32, 20133 Milan, Italy}

\title{On-chip Brillouin notch filter on a silicon nitride ring resonator}


\begin{document}


%\begin{tocentry}
%\centering
% \includegraphics{Figures/Fig0.pdf}
%\end{tocentry}

\begin{abstract}

Noncontact Brillouin spectroscopy is a purely optical and label-free method to retrieve fundamental material viscoelastic properties. Recently, the extension to a three-dimensional imaging modality has paved the way to novel exciting opportunities in the biomedical field, yet the detection of the Brillouin spectrum remains challenging as a consequence of the dominant elastic background light that typically overwhelms the inelastic Brillouin peaks. In this Letter, we demonstrate a fully integrated and ultra-compact Brillouin notch filter based on an optical ring resonator fabricated on a silicon nitride platform. Our on-chip ring resonator filter was measured to have a $\sim$10$\,$dB extinction ratio and a Q factor of $\sim1.9\cdot10^5$. The experimental results provide a proof-of-concept on the ability of the on-chip filter to attenuate the elastic background light, heralding future developments of fully integrated, ultra-compact and cost-effective Brillouin spectrometers. 

%This is 1-2 orders of magnitude lower than typically achieved with chip-scale low-cost sensors.

\end{abstract}

 Characterization of the material mechanical properties is a primary task across different fields ranging from chemistry to material and life sciences. Unlike standard analytical approaches based on the application of contact forces, Brillouin spectroscopy is purely optical and provides access to 3D mechanical properties without the need of contact nor sample labeling \cite{prevedel2019brillouin}. In Brillouin spectroscopy, a narrowband monochromatic laser beam illuminates the sample and the scattered light is spectrally analyzed by a spectrometer with sub-GHz resolution \cite{palombo2019brillouin}. The frequency shift and the linewidth of the Brillouin spectral peaks arising from the light interaction with thermally-activated spontaneous acoustic waves of matter provides information about the individual elastic moduli that form the full material elastic tensor \cite{koski2013non}. The recent combination of Brillouin spectroscopy with confocal microscopy \cite{koski2005brillouin, scarcelli2008confocal} has enabled a wide variety of applications within the life sciences including the elasticity assessment of the ocular lens and cornea \textit{in-vivo} \cite{scarcelli2012vivo}, the 3D mechanical imaging of living cells \cite{scarcelli2015noncontact, antonacci2016biomechanics, elsayad2016mapping, coppola2019quantifying} as well as the investigation of the liquid-to-solid phase transitions of intracellular compartments \cite{antonacci2018background, schlussler2021combined, de2018mutant}. Moreover, Brillouin spectroscopy holds promises to become a  diagnostic tool in clinical environments \cite{rioboo2021brillouin}, where altered biomechanical properties are understood to be the primary initiators of age-linked pathologies such as keratoconus \cite{scarcelli2014biomechanical}, cancer \cite{margueritat2019high, troyanova2019differentiating} and atherosclerosis \cite{antonacci2015quantification}. 

Despite the broad range of potential applications \cite{antonacci2020recent},  Brillouin spectroscopy remains challenging in the spontaneous regime as a consequence of the dominant elastic background light arising from both Rayleigh scattering and specular reflections. When the relative strength difference between the elastic and inelastic spectral components exceeds the contrast of the spectrometer, the Brillouin peaks are overwhelmed and cannot be detected. Multipass Fabry-Perot interferometers of high ($\sim$10$^{15}$) spectral contrast have for long been used in Brillouin spectroscopy \cite{scarponi2017high}, but their relative slow scanning process imposes long (typically $>1\,$sec) data acquisition time that makes them not suitable for imaging. To overcome this limit, a modified non-scanning version of the solid Fabry-Perot etalon, namely a Virtually Imaged Phased Array (VIPA), has been used in Brillouin microscopy \cite{scarcelli2011multistage}. Yet, its limited ($\sim$30$\,$dB) spectral contrast imposes complex multistage architectures that affect the system throughput efficiency and increase complexity. Extensive research has been carried out both to increase the spectral contrast of VIPA-based spectrometers and to suppress the unwanted elastic background light.  Beam apodization have been used to equalize the exponentially decaying beam at the output of the VIPA using linear absorption filter \cite{scarcelli2015noncontact} and spatial light modulators  \cite{antonacci2016breaking}. Other successful methods for spectral contrast enhancement involved the use of diffraction masks deflecting the background light from the dispersion axis  \cite{antonacci2018background} and Lyot filters removing the high-frequency content of the spectrum \cite{edrei2017integration}. On the other hand, suppression methods involved the use of cells containing media absorbing at the laser wavelength \cite{meng2014background}, dark-field illumination using annular beams \cite{antonacci2017dark} and different interferometric mechanisms aiming at the destructive interference of the elastic background light \cite{antonacci2015elastic, fiore2016high, shao2016etalon, lepert2016assessing}. While these methods offer a spectral visibility gain on the order of 30-40 dB, they rely on bulk optomechanics that generally affect the system robustness and increase complexity.

\begin{figure}[h!]
\centering\includegraphics[width=0.7\linewidth]{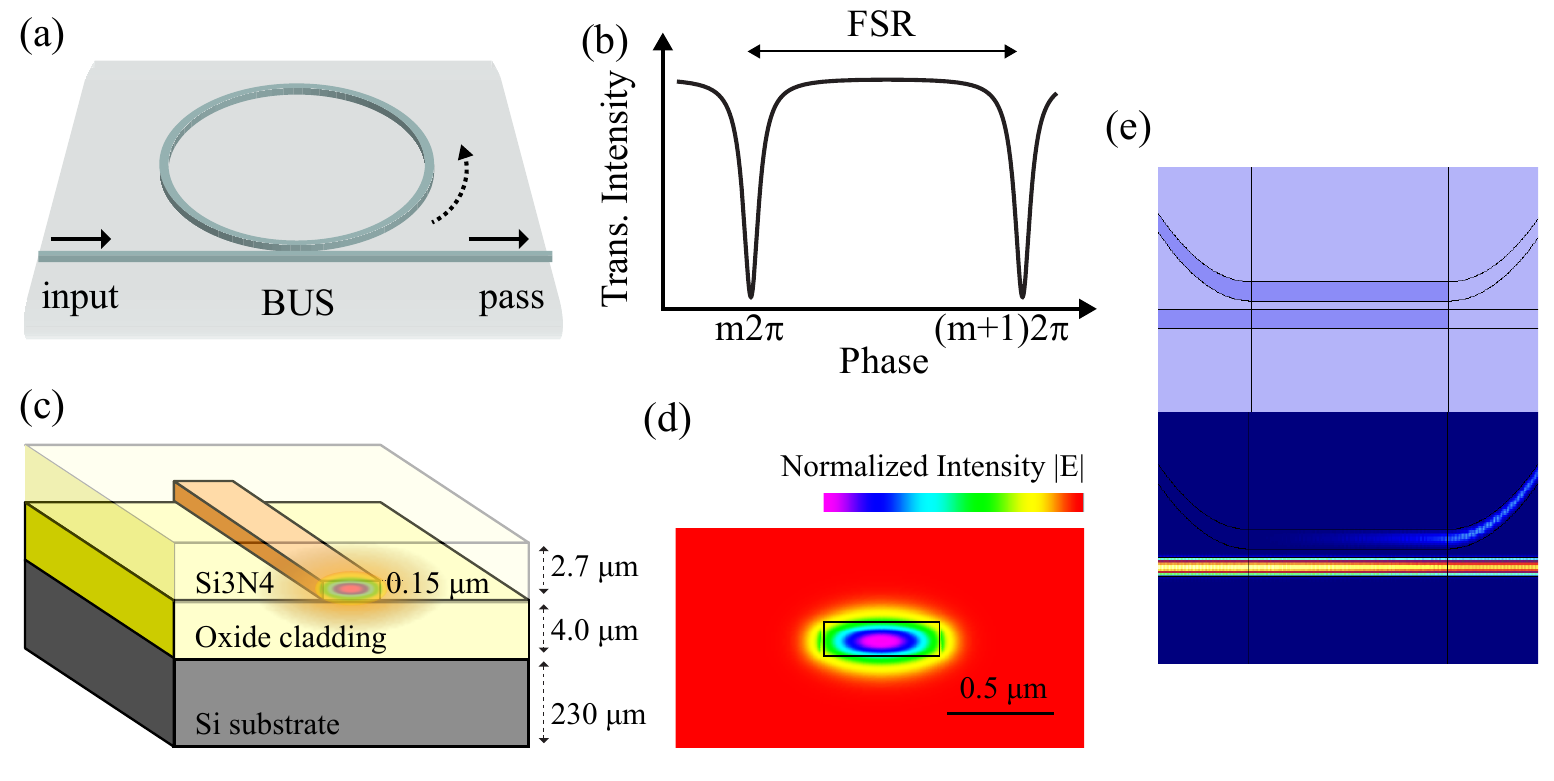}
\caption{All-pass ring resonator (a). Light is coupled to the ring from a BUS waveguide. The transmission intensity profile at the pass port of the ring exhibits sharp periodic dips at the ring resonances occurring at each $2\pi$ roundtrip phase shifts (b). PIC structure (c). The Si$_3$N$_4$ waveguides of $0.15 \times 0.60\, \mu$m$^2$ are embedded in a SiO$_2$ cladding, providing tight confinement of the fundamental TE mode (d). Simulation of the ring coupling in a racetrack configuration (e). }
\label{Fig1}
\end{figure}
In this Letter, we demonstrate  an integrated microring resonator working as a notch filter to suppress the elastic background light. A ring resonator is a resonant cavity integrated on a Silicon Photonics chip where the resonance condition occurs when the optical length of the ring is exactly an integer number  $m$ of the central wavelength (Fig. \ref{Fig1}a) \cite{bogaerts2012silicon}. High Q factor ring resonators have been demonstrated and are extensively used for a broad range of applications including datacom \cite{oda1988wide} and sensing \cite{sun2011optical}. The recent development of the high index contrast silicon nitride (Si$_3$N$_4$) platform has extended the use of Photonics Integrated Circuits (PICs) for the visible wavelengths \cite{rahim2017expanding}, in turn heralding opportunities in fields such as spectroscopy \cite{dhakal2014evanescent}, sensing \cite{antonacci2020ultra} and microscopy \cite{tinguely2017silicon}. 

We exploited the Si$_3$N$_4$ platform to design and fabricate a narrowband ring-based Brillouin notch filter for operation at $\lambda=532\,$nm central wavelength. The ring was conceived in an all-pass mode so as to provide a similar transfer function to that of a Fabry-Perot interferometer working in reflection, namely an almost flat transmission intensity with narrow dips occurring at the ring resonance and equally spaced in frequency (Fig. \ref{Fig1}b). The frequency separation of two consecutive resonance peaks is defined by the Free Spectral Range (FSR) given by the expression
\begin{equation}
\text{FSR} = \frac{\lambda^2}{n_gL},
\label{eq1}
\end{equation}
where $n_g$ is the group index and $L$ the ring roundtrip length (Fig. \ref{Fig1}b). From Eq. \ref{eq1}, we set a ring radius of $R=565\, \mu$m to have a FSR$=40\,$GHz. The PIC was fabricated through a low pressure chemical vapor deposition (LPCVD) process for an accurate deposition of the silicon oxide and silicon nitride layers. A representative layout of the PIC layer structure is shown in Fig. \ref{Fig1}c. The resulting Si$_3$N$_4$ waveguides had a nominal dimension of $x=0.60\,\mu$m and $y=0.15\,\mu$m to ensure a tight confinement of the fundamental TE mode and thus to minimize the bending losses. Fig. \ref{Fig1}d shows the simulated TE mode with respect to the Si$_3$N$_4$ waveguide. To finely tune the resonant wavelengths of the ring, a phase shifter is needed to change the effective refractive index $n_{eff}$ of the ring waveguide. In the present study, a $2\pi$ phase shift was achieved by locally heating the Si$_3$N$_4$ waveguide of the ring through the application of an electrical current to a metal resistor placed above the ring. To maximize the extinction ratio of the filter, the ring was designed in an attempt to achieve a critical coupling, a condition that occurs when the coupled power equals the power loss of the ring. Assuming an estimated attenuation coefficient of $\alpha=1\,$dB/cm, this condition is satisfied for a ring coupling power efficiency of $k=12.2\%$. To this aim, the ring was designed with a minimum gap of $275\,$nm and a racetrack configuration, i.e. with an elongated shape of $93.86\,\mu$m along the direction of the BUS waveguide (Fig. \ref{Fig1}e). The racetrack configuration was needed to compensate for the tight confinement of the evanescent field across the waveguides and the minimum gap size imposed by the fab resolution in an attempt of achieving the target coupling power.

\begin{figure}[h!]
\centering\includegraphics[width=0.7\linewidth]{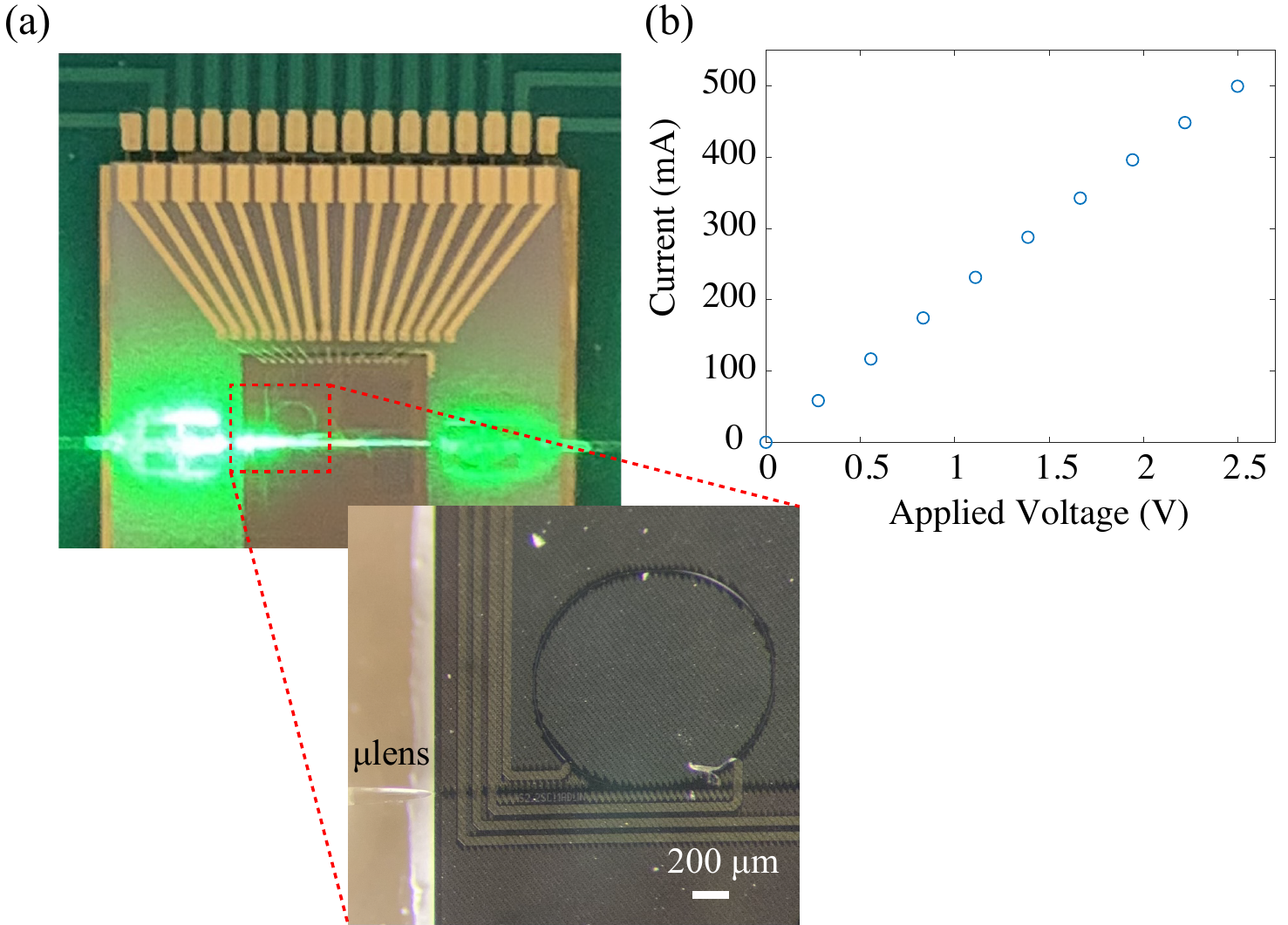}
\caption{Functional PIC assembly (a). The optical coupling was achieved using microlensed single-mode fibers that partially mitigated the MFD mismatch with the Si$_3$N$_4$ waveguides. The PIC was mounted on a PCB to ease electrical accessibility. Current plot as a function of the applied voltage, indicating a ring heather resistance of $\sim5\,\Omega$ (b).}
\label{Fig2}
\end{figure}
To functionalize the filter, the PIC was packaged so as to provide both electrical and optical coupling (Fig. \ref{Fig2}a). To ease the electrical accessibility needed for the thermal tuning, the PIC was mounted and wire bonded on a custom printed circuit board (PCB), whose inputs and outputs were connected to a current supplier (Keysight E36312A). Preliminary tests were conducted to verify  the electrical conductivity of the heaters and its resistance by reading the circuit current in response to an applied voltage (Fig. \ref{Fig2}b). Given the mode field diameter (MFD) mismatch between a conventional single-mode fiber for visible light ($\text{MFD}\sim 4 \mu$m) and our Si$_3$N$_4$ waveguides ($\text{MFD}_x\sim 0.50 \mu$m, $\text{MFD}_y\sim 0.27 \mu$m), a direct fiber-to-chip coupling would result in high ($>40\,$dB) insertion losses that may prevent the detection of the Brillouin signal. To overcome this limitation, we employed two lensed single-mode fibers (Thorlabs SM450) of NA$\sim0.4$ and nominal $\text{MFD}=0.8\, \mu$m in an attempt to limit the optical insertion loss, which we measured to be $\sim18\,$ dB per facet. Although around $5\,$dB insertion loss per facet may be achieved in principle with the present configuration \cite{sui2008automatic}, we obtained higher loss as a consequence of the challenging alignment process given by the nm accuracy requirement as well as the micro-mechanical drifts arising during the epoxy curing process.

\begin{figure}[h!]
\centering\includegraphics[width=0.6\linewidth]{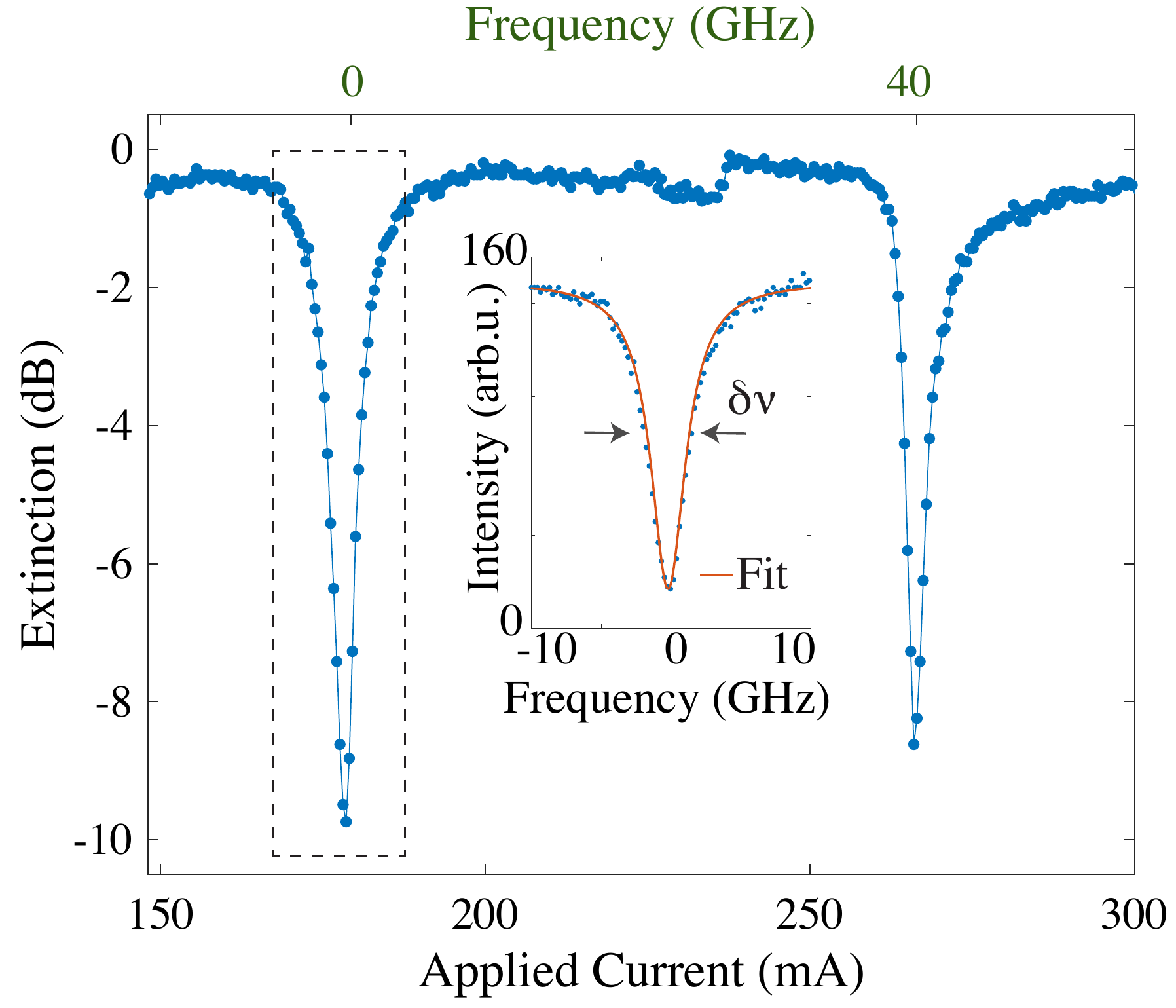}
\caption{Spectral profile (in dB units) of the ring resonator for monochromatic ($\lambda=532\,$nm) light and for two consecutive ring resonances as a function of both frequency (top axis) and applied current  (bottom axis). Resonant peak profile (in linear scale) and associated Lorentzian fit as a function of frequency (inset). Data show an extinction ratio of $9.7\,$dB and a peak FWHM $\delta \nu=3.0\,$GHz, corresponding to a ring Q factor of $1.9\cdot10^5$. }
\label{Fig3}
\end{figure}
We measured the extinction ratio of our on-chip ring filter by coupling the monochromatic light of a narrowband single longitudinal mode laser with high spectral purity (Laser Quantum torus, $\lambda=532\,$nm) directly to the PIC and finely tuning the ring for two consecutive interference orders, i.e. for a round trip phase change of  $2\pi$. The spectral profile was then reconstructed by measuring the light signal transmitted trough the pass port of the ring at the output of the lensed single-mode fiber. Fig. \ref{Fig3} shows the resulting spectral profile of the ring for two consecutive resonances. By fitting the peaks with a Lorentzian function, we measured an extinction ratio of $\sim9.7\,$dB, a finesse $F = \text{FSR}/\delta \nu \sim 13.3$ and a peak linewidth of $\delta \nu \sim3\,$GHz, indicating a Q factor of $\text{Q}=\nu/\delta \nu \sim 1.9\cdot 10^{5}$ in line with our expectations.

\begin{figure}[h!]
\centering\includegraphics[width=0.8\linewidth]{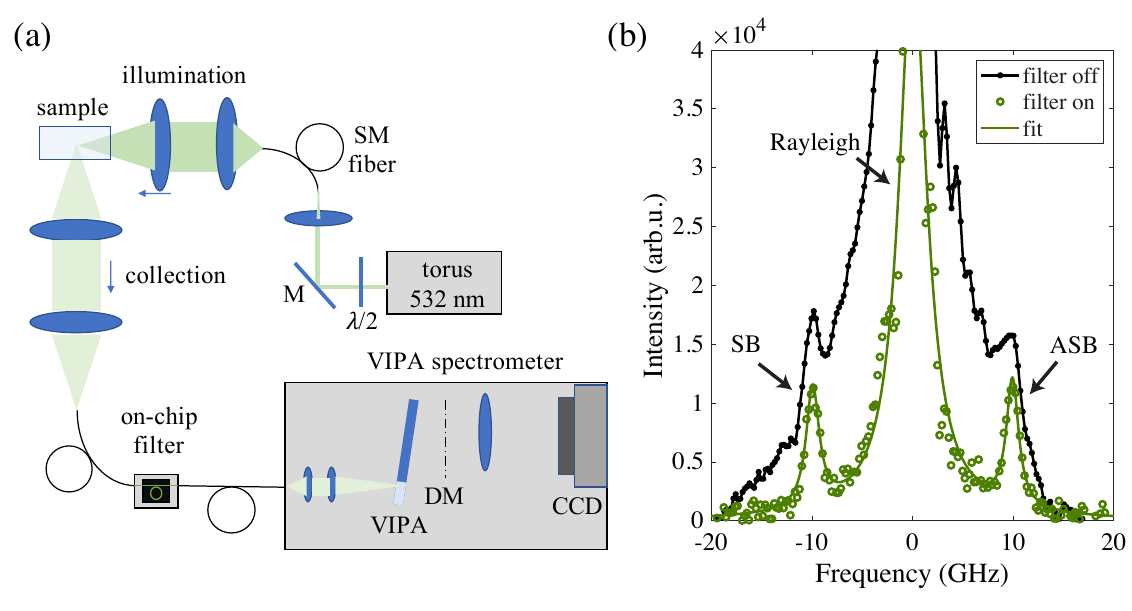}
\caption{Brillouin spectroscopy platform. The laser beam was coupled into a single-mode fiber (SM fiber) and focused to the sample. The light scattered is collected at $90^{\circ}$ by a single-mode fiber working as a confocal pinhole and coupled to the on-chip filter for elastic background cleaning. The output fiber of the filter is then connected to a single-stage VIPA spectrometer for parallel spectral acquisition. A CCD camera is used to detect the Stokes (SB) and Anti-Stokes (ASB) Brillouin peaks of polystyrene with the filter turned on and off (b).  $\lambda/2$ half wave plate; M mirror; DM diffraction mask.  }
\label{Fig4}
\end{figure}
A Brillouin optical system (Fig. \ref{Fig4}a) was built in a $90^{\circ}$ scattering geometry to provide a proof-of-concept experiment of the capability of the on-chip ring filter to suppress the elastic background light in Brillouin spectral measurements. The laser beam was first coupled into a single-mode fiber for mode cleaning and then focused to a polystyrene cuboid of transparent and polished facets that was used as a test sample. The scattered light was collected at $90^{\circ}$ by a lens of equal $\text{NA}=0.2$ with respect to the illumination one and delivered to our on-chip filter thought the assembled fiber, which also acted as a confocal pinhole at the collection arm. The output fiber of the PIC was then connected to a single-stage VIPA spectrometer integrating a VIPA etalon (LightMachinery) of FSR$=60\,$GHz, a diffraction mask for spectral contrast enhancement \cite{antonacci2018background} and a CCD camera (FLIR BFS-U3-63S4M-C). 

Fig. \ref{Fig4}b shows the Brillouin spectrum of the polystyrene test sample acquired by the VIPA spectrometer before and after tuning the ring resonance at the laser wavelength. Given the relatively high optical insertion loss at the PIC input and output waveguides, we set an illumination power of $\sim150\,$mW at the sample plane and a data acquisition time of 30 sec averaging over 5 repetitions. With the ring off resonance, the Brillouin peaks were almost completely dominated by the elastic crosstalk light arising from the amorphous structure of polystyrene. In turn, the visibility of the Brillouin peaks increased significantly as the ring resonance was finely tuned at the laser wavelength. Fitting the Brillouin peaks with a Lorenzian function, we measured a Brillouin shift of $\nu_B = 9.9\,$GHz and the acoustic velocity $V=2330\,$m/s, in good agreement with the expected values for polystyrene \cite{forrest1998brillouin}.

In conclusion, we demonstrated a fully integrated on-chip Brillouin notch filter based on a racetrack all-pass ring resonator fabricated on a silicon nitride platform. While racetrack ring resonators are common components in integrated photonics, their application as notch filters for Brillouin spectroscopy in the visible domain represents a key cornerstone in an attempt of eliminating the need of free-space optics and developing ultracompact and miniaturized systems. 
%Our ring has a measured extinction ratio of $9.7\,$dB and a Q factor of $1.9\cdot10^5$. 
Although the measured extinction ratio achieved in the present study is not yet comparable with the $30\,$dB typically obtained by standard Brillouin bulk filters, our integrated notch filter provides unprecedented robustness, ease-of-use and stability as neither free-space optical components nor moving parts are needed. Higher extinction may be reached depending on the level of balance  between the ring coupling power and the roundtrip loss. Nevertheless, an exact knowledge of the coupling as a function of the ring gap size and racetrack length is non-trivial as a consequence of the fabrication tolerances and typically requires multiple fabrication runs. On the other hand, higher extinction ratios may also be achieved by cascading multiple ring resonators in tandem with no significant increase in the system footprint, complexity nor insertion loss. Besides the increase in the extinction ratio, future implementations will also involve a significant reduction in the PIC insertion loss, which may extend the applicability of the on-chip notch filter to Brillouin microscopy where a high throughput efficiency is needed to enable a fast  ($<100\,$ms) data acquisition with minimal ($<10\,$mW) optical power for probing living biological systems. Results pave the way to further development of on-chip notch filters not only  for Brillouin spectroscopy, but also for the other spectroscopy methods working with visible light such as low-frequency Raman. Last but not least, improvements in the ring Q factor at visible wavelengths may also inspire future implementations of ultracompact and cost-effective spectrometers.

\begin{acknowledgement}

We thank Dr. Gregor Klatt and Laser Quantum - A Novanta Company for a free loan of \textit{torus} SLM laser.

\end{acknowledgement}

%\section{Disclosures}
%GA and DP own share of Specto S.r.l.

\bibliography{biblio.bib}


%\begin{suppinfo}
%\end{suppinfo}


\end{document}